\documentclass[aps,prb,twocolumn,showpacs,a4paper,unsortedaddress,preprintnumbers,amsmath,amssymb,byrevtex]{revtex4}

\usepackage{graphicx}
\usepackage{dcolumn}
\usepackage{bm}
\usepackage{hyperref}

\def\rr{\vec{r}}

\def\n{n(\vec{r})}
\def\nn{\tilde{n}(\vec{r})}
\def\m{\vec{m}(\vec{r})}

\begin{document}

\title{First principles simulations of the magnetic and structural properties of Iron}

\author{V. M. Garc\'{\i}a-Su\'arez}
\affiliation{Departamento de F\'{\i}sica, Universidad de Oviedo, 33007
Oviedo, Spain}
\email{victor@condmat.uniovi.es}
\author{C. M. Newman}
\author{C. J. Lambert}
\affiliation{Department of Physics, Lancaster University, Lancaster, LA1 4YB, U. K.}
\author{J. M. Pruneda}
\author{J. Ferrer}
\affiliation{Departamento de F\'{\i}sica, Universidad de Oviedo, 33007
Oviedo, Spain}
\email{ferrer@condmat.uniovi.es}


\date{\today}

\begin{abstract}
We have implemented non-collinear GGA and a generalized Bloch's theorem to simulate
unconmensurate spiral arrangements of spins in a Density Functional Theory code based
on localized wave functions. We have subsequently performed a thorough study of the
different states of bulk Iron. We determine the minimal basis set required to obtain
reliable orderings of ground and excited states. We find that the most stable fcc
phase is a spiral with an equilibrium lattice constant 3.56 \AA.
\end{abstract}


\maketitle

\section{introduction\label{introduction}}
Advances in experimental setups along the past two decades have allowed to
grow in a controlled way, and characterize, tiny structures and devices, 
paving the way for the slow development of those fields of Materials Science
now covered under the umbrella of Nanoscience. Ever more powerful computers 
and canny algorithms have also allowed to simulate larger and larger clusters
of atoms or molecules, filling partially the bridge between theory and experiments.

Molecular Dynamics packages based on Density Functional Theory (DFT)\cite{Koh65} 
represent a specially useful set of theoretical tools in the analysis of such materials 
and devices. Among them, those which use basis sets (BS) of localized wave functions are 
particularly attractive since, one the one hand, can be easily adapted to work with order-N 
algorithms, and on the other, they can be written in a tight-binding language, which allows 
for easier analytical approaches. 

SIESTA is a simulation package which literally implements the tight-binding 
philosophy. \cite{Sol02} Indeed, it uses Norm-Conserving 
pseudopotentials \cite{Tro91} to integrate away core energy levels and very flexible 
BS made up of numerical atomic-like wave-functions to handle valence electrons. 
The use of minimal bases allow for fast calculations which already provide a 
qualitative understanding of the simulated material. Accurate calculations
can be performed at a  higher numerical cost, using more complete BS. Assessment of the
degree of reliability of a BS might be essential, since competing would-be
ground states may in some instances have small energy differences. Such analyses have 
already been performed for selected molecules and solids \cite{Jun01,Ang02}, but not for 
magnetic materials. Those studies show how both the number of wave functions
used as well as their extent are variational parameters, providing therefore a path for
systematic improvements of the accuracy of a simulation. A similar study for magnetic
elements or materials seems to be highly desirable, since they have their own peculiarities and,
in particular, are usually tougher to simulate. We have performed an 
exhaustive study of the degree of accuracy of the basis for Iron in most of its
bulk phases as well as for small clusters. We find that SIESTA provides a highly  accurate 
description of the systems we have scrutinized, provided that a large numbers of extended 
orbitals is used. We will show below how BS regarded as fairly complete for simpler elements
can provide disastrous results for this transition metal.
 
The SIESTA package has built into it the possibility to cope with non-collinear
commensurate spin structures but only in the Local Density Approximation (LDA).
We have therefore coded the ability to compute non-collinear arrangements of the spin 
moments in the Generalized Gradient Approximation (GGA), since LDA fails to provide adequate
ground states and lattice constants of a number of magnetic transition metals. Moreover, we
have included the possibility to simulate non commensurate spiral structures \cite{San98,Sti89}.

While bcc ($\alpha-$)Iron is firmly established to be a ferromagnet, the scenario for 
fcc ($\gamma-$)Iron is more complex, since it stands at a crossing point between  
high spin ferromagnetic and antiferromagnetic states, and the actual realization
depends sensitively on its actual atomic volume and, possibly, strains \cite{Spi02}.  
Tsunoda discovered a decade ago that $\gamma-$Iron could be stabilized as pellets of
radii up to 100 nm, with a lattice  constant of 3.577 \AA \cite{Tsu89,Tsu93}. He also
found that the magnetic structure of the pellets was helicoidal, with pitch vector
$\vec{q}_{exp}=(0.12,0,1)$. A number of authors have subsequently looked for
theoretical low-energy collinear and non-collinear states appearing in such $\gamma-$phase.
\cite{Kor96,Mor97,Kno00,Mar02,Uhl92,Uhl94,Byl98,Byl99a,Byl99b}.

Kn\"opfle and co-workers \cite{Kno00}, who used GGA or LDA and a full-potential 
implementation of Density Functional Theory found that the ground state
was indeed a spiral with the correct pitch vector. But since the Augmented
Spherical Wave (ASW) method tends to overestimate the atomic volume, it is difficult
to extract what their equilibrium lattice constant $a_0$ might be. 

We have found a spiral ground state with $a_0$ of 3.56 \AA, in excellent agreement with 
the experimental data of Tsunoda.  We find two local minima when we plot the energy of
the spiral state as a function of the pitch vector, at $\vec{q}_1=(0,0,0.6)$ and
$\vec{q}_2=(0.12,0,1)$, as other authors did \cite{Uhl92}. We label these two states
as S1 and S2. The state S1 is the global minimum for lattice constants down to 3.47
\AA, and only below it is S2 the ground state. 

We believe that our results represent a significant methodological advance for the 
simulations of magnetic systems using the SIESTA package since, on the
one hand, we lay down a firm ground for the reliability of atomic bases of different 
sizes and, on the other, we allow for the description of interesting non-collinear and
spiral structures of the spin.

The layout of this article is simple. Section II provides the theoretical backbone of
the article; Subsection IIa is a brief reminder of the non-collinear formalism as is
applied to DFT; we present subsequently details of our implementation of 
non-collinearity and non commensurate spiral arrangements of spin for such structures.
Section III is devoted to show and discuss our results for the stability of the different
states of Iron, both within LDA and GGA, using different BS, up to an optimal choice.
We finish the article with a short summary.

\section{Theoretical backbone}
We devote this section to provide details of our implementation of non-collinear GGA and 
of unconmensurate spiral arrangements. We believe it is useful to supply first a backbone of 
non-collinear DFT \cite{Kub88,San98}, which will help us discuss similarities and
differences with the latest approaches \cite{Kno00}.

\subsection{Brief presentation of non-collinearity in DFT}   
\begin{enumerate}
\item In a non-collinear material, the direction of the magnetization vector $\m=m\cdot\vec{u}_m$ 
changes at each place in the sample, according to the angles $\theta$, $\phi$, 

\begin{equation}
\vec{u}_m=(\,sin(\theta)\,cos(\phi),sin(\theta)\,sin(\phi),cos(\theta)\,) 
\end{equation}

\item The density matrix can be decomposed in terms of the electronic density $\n$ and $\m$ as 

\begin{equation}
\nn=\frac{1}{2}\,(\,\n+\m\cdot\tilde{\vec{\tau}}\,)
\end{equation}

where $\tilde{\vec{\tau}}$ denote the three Pauli matrices. There is a single rotation matrix,
$\tilde{U}(\theta(\rr),\phi(\rr))=e^{i\tilde{\tau}_y\theta/2}\,
e^{i\tilde{\tau}_z\phi/2}$
which brings $\nn$ to collinear form, 

\begin{eqnarray}
\left(\begin{array}{cc}n_{11} &n_{12}\\n_{21} &n_{22}\end{array} \right)= 
\hat{U}^{\dagger}\left(\begin{array}{cc}n_{\uparrow} &0\\0 &n_{\downarrow}
\end{array}\right)\hat{U}
\end{eqnarray}

\item The Total Energy is a functional of the density matrix, $E[\nn]=T[\nn]+E_H[\nn]+E_{xc}[\nn]$
which, upon variation provides with the effective single-particle Hamiltonian

\begin{eqnarray}
\tilde{H}_{DFT}=\left(-\frac{\hbar^2}{2\,m}\nabla^2+v_H[\nn]\right)\,\tilde{I}+v_{xc}[\nn].
\end{eqnarray}

$T$ and $E_H$ are the kinetic and Hartree energy functionals, while

\begin{equation}
E_{xc}= \int d\rr f_{xc}(n,m,\vec{u}_m,\nabla n,\nabla m,\nabla\vec{u}_m)
\end{equation}

takes account of exchange and correlation. $v_{H}$ and $v_{xc}$, in Eq. (4), are the
corresponding potentials.

\item The spinor eigenfunctions of $\tilde{H}$, $\tilde{\psi}_i(\rr)$ can be used
to compute the density matrix, since $\nn=\sum_i \tilde{\psi}_i\,\,\tilde{\psi}_i^\dagger$ 
Each eigenfunction can individually be rotated to bring it back to collinear form

\begin{eqnarray}
\tilde{\psi}_i\rr)=\left(\begin{array}{c}\psi_1(\rr)\\\psi_2(\rr)\end{array}\right)=
\tilde{U}_i(\rr)\left(\begin{array}{c}\phi(\rr)\\0\end{array}\right)
\end{eqnarray}

\item The exchange and correlation potential matrix, which is obtained by functional
differentiation of the exchange and correlation energy, can be uniquely decomposed in terms
of Pauli matrices,

\begin{equation}
\tilde{v}_{xc}=\frac{\delta E_{xc}}{\delta \nn}=v_s \,\tilde{I}+\vec{v}_v \cdot\tilde{\vec{\tau}}
\end{equation}

where $\vec{v}_{s}={\tt tr}(\tilde{v}_{xc}\tilde{I})/2$ and 
$\vec{v}_{v}={\tt tr}(\tilde{v}_{xc}\tilde{\vec{\tau}})/2$.

\item In LDA approximation, $\vec{v}_v$ is a function of only one vector, $\vec{u}_m$ 
so that it must be proportional to it. It is then 
easily shown that
$\tilde{v}_{xc}$ is diagonalized by the same rotation matrix as $\tilde{n}$,

\begin{eqnarray}
\tilde{v}_{xc}=\tilde{U}^{\dagger}\,\tilde{v}_{xc}^{coll}\,\tilde{U}=
\tilde{U}^{\dagger}\left(\begin{array}{cc}v_{\uparrow}&0\\
0&v_{\downarrow}\end{array}\right)\tilde{U}
\end{eqnarray}

where $v_{\sigma}=v_{\sigma}\left[n_{\sigma}\right]$.
This can be interpreted physically as rotating the whole system into a collinear
reference frame, where $v_{\sigma}(\rr)$ can be computed as in conventional LDA.
\item  The GGA expression for the exchange and correlation energy contains also the
vector $\nabla^2 \vec{u}_m$. Therefore the GGA potential includes both spin stiffness
and antisymmetric exchange terms (Dzyaloshinskii-Moriya \cite{Dzy58,Mor60})

\begin{equation}
\vec{v}_{xc} = v_s\tilde{I}+(v_m\vec{u}_m+v_{grad}\nabla^2 \vec{u}_m+v_{cross}
\vec{u}_m\times\nabla^2\vec{u}_m)\tilde{\vec{\tau}}
\end{equation}

This implies that 
$\hat{v}_{xc}$ can not be fully diagonalized by the $\hat{U}$ rotation matrices,

\begin{equation}
\tilde{U}\hat{v}_{xc}\tilde{U}^{\dagger}=
\tilde{v}_{xc}^{coll}+\tilde{U}(v_{grad}\nabla^2\vec{u}_m
+v_{cross}\vec{u}_m\times\nabla^{2}\vec{u}_m)\tilde{\vec{\tau}}
\tilde{U}^{\dagger}
\end{equation}

\end{enumerate}

\begin{figure}
\includegraphics[width=\columnwidth]{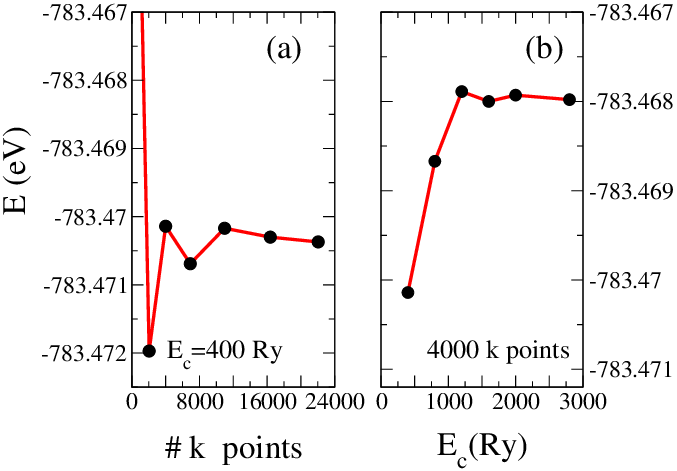}
\caption{\label{kpoints} Total Free Energy as a function of (a) number of k points in half the 
Brillouin zone (with a Grid cutoff of 400 Ry) and (b) Grid cutoff (with 4.000 k points in half the 
Brillouin zone). An optimized Double-Zeta (DZ) basis was used.}
\end{figure}

An accurate description of non-collinearity must include a spin stiffness
term \cite{Byl99a}, as is the case for classical localized spins. For instance,
the Heisenberg Hamiltonian is 

\begin{equation}
E \sim \rho \,m^2 \int d\rr (\nabla \vec{u}_m(\rr))^2,
\end{equation} 

\noindent where $\rho$ is the spin stiffness.

Our recipe to partly take account of gradient corrections is to neglect 
the stiffness contribution and evaluate only $\tilde{v}_{xc}^{coll}$. 
We rotate the density matrix to bring it into collinear form, $\tilde{n}^{coll}$.
We then compute its gradients $\nabla\tilde{n}^{coll}$ in this collinear reference
frame, and the collinear potential, $\tilde{v}^{coll}$. We finally rotate back. 
Kn\"opfle and coworkers chose an apparently
different approach \cite{Kno00}. They rotated both $\nn$ and its gradient, 
$\nabla\nn$, and then evaluated $v_{xc}$ with the diagonal terms of both
matrices, discarding the non-diagonal terms of $\nabla\nn$. But since 

\begin{eqnarray}
\tilde{U}\nabla\tilde{n}\tilde{U}^{\dagger}&=&\nabla\tilde{n}^{coll}+
\tilde{U}(\nabla\tilde{U}^{\dagger})\tilde{n}^{coll}+\tilde{n}^{coll}
(\nabla\tilde{U})\tilde{U}^{\dagger}\\&=&\nabla\tilde{n}^{coll}+
\vec{A}\cdot\tilde{\vec{\tau}}
\end{eqnarray}

\noindent where $\vec{A}=(a_{x},a_{y},0)$, both approaches are analytically equivalent.
Kleinman and Bylander \cite{Kle99,Byl99b} added a spin stiffness
term to their LDA exchange and correlation potential from which they obtained a spiral ground
state with $\vec{q}=(1,0,0)$ at the atomic volume of bulk copper.

The effective LDA Hamiltonian does not commute with the Pauli matrices $\tilde{\tau}_{x,y}$
unless the system be paramagnetic or collinear ($\theta$ be equal to $0^o$). Therefore the
expectation values of $S_{x,y}$ are not conserved in the iterative selfconsistency process
of DFT. States with $\theta=90^0$ can be shown to be metastable, and therefore $\theta$ is
conserved in this case.

\subsection{Description of Non-collinear commensurate and spiral states for a 
localized BS} 
A convenient variational wave function for either molecules (or solids) with non-collinear 
(commensurate) magnetic moments is

\begin{equation}
\tilde{\psi}_{\alpha}(\rr)=\sum_{i} \phi_i(\rr-\vec{R_i})
\left(\begin{array}{c}c_{\alpha,i,1}\\c_{\alpha,i,2}\end{array}\right)
\end{equation}

For solids, the above wave function can be easily rewritten so that it explicitly satisfies
Block theorem.
For helicoidal arrangement of spins of pitch vector $\vec{q}$, the DFT Hamiltonian commutes with the
operator $T(\vec{R},\vec{q})=\tilde{U}(0,\vec{q}\cdot\vec{R})\,T(R)$, which translates by a 
lattice vector $\vec{R}$ and then rotates about the z-axis. Since a wave function of the kind    

\begin{equation}
\tilde{\psi}_{\vec{k}}^{\vec{q}}(\rr)=\sum_{\vec{R},i} e^{-i\vec{k}(\vec{R}+\vec{d}_i)}
\,\phi_i(\rr-\vec{R}-\vec{d}_i) \,\,\tilde{U}^{\dagger}(\theta_0,\vec{q}\cdot\vec{R})
\left(\begin{array}{c}c_{\vec{k},i,1}\\c_{\vec{k},i,2}\end{array}\right)
\end{equation}

\noindent is an eigenfunction of $T(\vec{R},\vec{q})$, a generalized Bloch theorem holds. \cite{San98}
It must be stressed that such wave function is not an eigenstate of $H_{DFT}$ since, as noted
above, a rotation by a constant angle $\theta_0$ about the y-axis, $\tilde{U}(\theta_0,0)$,
does not commute with the Hamiltonian, unless $\theta_0=0$. We have checked numerically
that for any wave-function of the form above, the angle $\theta_0$ is indeed not conserved 
by the application of $H_{DFT}$, unless $\theta=90^0$, which corresponds to a metastable 
situation. Such wave function must therefore be regarded as purely variational.

\begin{figure}
\includegraphics[width=\columnwidth]{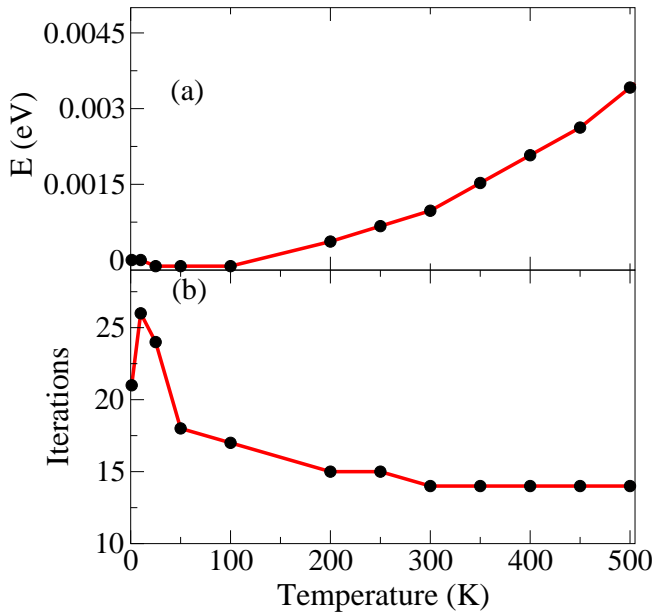}
\caption{\label{temp} Energy difference (a) and number of iterations (b) as a function of
the electronic temperature. The calculation was performed with 4.000 k points, a grid
cutoff of 400 Ry (27.000 points) and an optimized DZ basis. }
\end{figure}

\section{Results}
\subsection{Preliminaries: choice of pseudopotential and integration grids}
SIESTA uses norm-conserving pseudopotentials (NCPS) \cite{Tro91} optimized so that their 
local part be smooth \cite{Sol02}. Izquierdo and coworkers \cite{Izq00} proposed to 
generate the NCPS for Iron from the atomic configuration [Ar]3d$^7$4s$^1$, with core
radii for 4s, 4p, 3d and 4f orbitals set equal to 2.00 a.u. They found that the optimal
radius for partial-core corrections was 0.7 a.u.
We have looked carefully for a pseudopotential which could produce a better fit to bulk
bcc Iron. Our first criterium to determine the cutoff radii was to compare the eigenvalues 
of the valence shells of atomic Iron obtained from the NCPS and the all-electrons 
Hamiltonians and try to minimize their differences. We found that the radii obtained using
such procedure were very different from each other. Moreover, they produced 
fits to the bulk bcc phase of poor quality as compared with the proposal of Izquierdo et 
al. Inclusion of the partial-core 3p levels into the valence did not help, mostly due to
the fact that 3s electrons are still taken as part of the core and therefore there remains
a strong overlap between valence and pseudo-core charge.

SIESTA performs Brillouin zone integrations on a grid of Monkhorst-Pack special points 
typically extended to cover half of it \cite{Sol02,Mor92}. The integrand is also smeared by 
a Fermi function. Hamiltonian matrix elements are 
partly computed on a real space grid, whose fineness $\Delta x$ is controlled by a grid 
cutoff, $E_c\approx (\pi/\Delta x)^2/2$. 

Since the energy of the different states need not shift rigidly when increasing accuracy and
moreover competing ground states for (fcc) Iron have energy differences as tiny as 
5 meV, we decided to set the number of k points, the electronic temperature and the
grid cutoff to match an accuracy of about 1 meV. Figs. \ref{kpoints} and 
\ref{temp} show typical 
results for the convergence of the energy of bcc ferromagnetic Iron as a function of 
those parameters. All the calculations shown there were performed using the GGA functional
as parametrized by Perdew and coworkers \cite{Per96} and optimized either Double-Zeta 
or Triple-Zeta bases (see below). We find then that we need 4.000 k points and up to 700 Ry 
(which corresponds to 50.000 points in the real space grid) to meet the desired accuracy.
Fig. \ref{temp} shows that increasing the temperature 
to an optimal value of about 200 or 300 K speeds the convergence of the selfconsistent 
process significantly while not damaging the accuracy required for the energy.  

\subsection{Optimization of the atomic basis and Phase diagram for bulk Iron}

\begin{figure}
\includegraphics[width=\columnwidth]{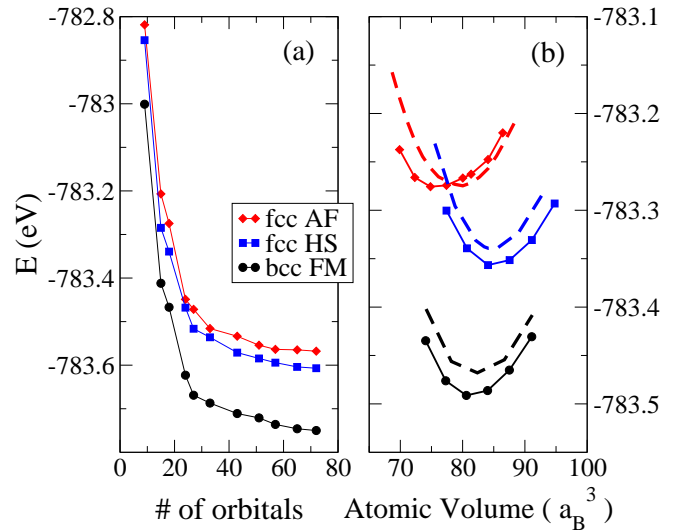}
\caption{\label{hilbert} (a) Evolution of the Free Energy of the three most stable states of Iron 
as a function of the size of the BS. AF = antiferromagnetic, HS = high-spin, FM = ferromagnetic.
(b) Cohesive energy curves of those same three states, for the two minima found using the BSD.
The discontinuous curves correspond to radii of 6 a. u. and the continuous to 10 a. u.The calculations
have been performed using the GGA approximation.}
\end{figure}

SIESTA allows for a great flexibility in the use of BS of wave functions which describe
valence electrons. For each species of atom, one may specify one shell of s, p, d and f 
orbitals. Within each shell, one may choose how many wave functions having
the required angular symmetry are needed. A Single-Zeta basis (SZ) is equivalent to 
choosing just one. Completion of the basis leads to Double-Zeta and Triple-Zeta bases 
(DZ, TZ). In addition, one may polarize an orbital (P), which means adding wave functions 
which correspond to one higher angular momentum unit \cite{Sol02}. The minimum basis 
required to accommodate the eight valence electrons of Iron would be SZ for both s and d 
orbitals, which provides a total of 6 wave functions per spin. SIESTA is
set up to the default maximum BS TZTP, which corresponds to 72 wave 
functions (WF) per spin. Using more wave is equivalent to filling up the Hilbert space and
provides a better variational estimate of the ground state. Junquera and coworkers 
\cite{Jun01} pointed out that the confinement radius of each orbital are also variational
parameters. Very fast calculations or simulations of a large number of atoms may therefore be
performed by the use of a Single-Zeta basis of rather confined orbitals. Such calculations 
usually provide much of the features of a material or device. But 
they are usually regarded are pretty inaccurate, and DZ bases with polarized s orbitals are
rather used. 

We have minimized a few BS ranging from SZ-SZ-SZ (9 WF) to TZTP-TZTP-TZTP (72 WF). 
Fig. \ref{hilbert} (a) shows how the convergence of the energy
for the three most stable states of bulk Iron, e. g.: bcc ferromagnetic, fcc ferromagnetic
high spin and fcc antiferromagnetic as a function of the number of orbitals used in the BS.
We find that a TZ-TZ-TZ BS (27 WF) is essentially converged for p, d and f orbitals, since the 
Free Energy of the three states changes only a little if we polarize this BS. 
While we have not checked explicitly that a fourth Zeta for the s-orbital may still change 
somewhat the energy, a inspection of the curve induces us to believe that our results are
completely converged. It is also apparent that the states do not shift rigidly upon 
improving the accuracy. We shall see below that such an effect is specially damaging in 
the LDA approximation.

\begin{figure}
\includegraphics[width=\columnwidth]{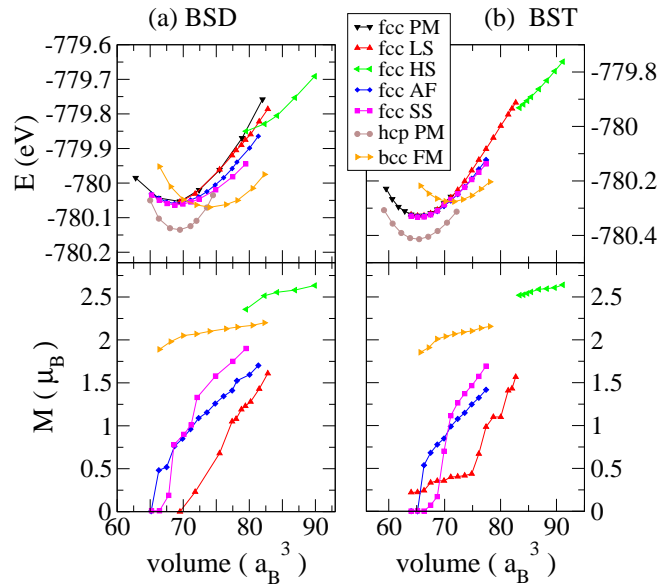}
\caption{\label{statesLDA} Free energy and magnetic moment of the ground and lowest excited
states of bulk Iron as predicted by LDA approximation, using (a) BSD and (b) BST, as
a function of the atomic volume.}
\end{figure}

We have paid special attention to the minimization of the BS DZP-SZ-DZ (BSD) and 
TZ-TZ-TZ (BST), where we have used a grid software program to look for local minima of the
energy as a function of the radii of the first Zeta of s, p and d orbitals. We find a first
local minimum for somewhat confined radii of about 6 a.u. and a deeper one for radii
of about 10 a.u. We found that the energy still decreased upon looking further away, but 
thought it worthless to attempt to look for such next minimum. Fig. \ref{hilbert} (b) shows
that extended radii improve both the energy and the lattice constant substantially.
For instance, the lattice constant of the bcc ferromagnetic state obtained using BSD
in GGA approximation, as predicted by the first minimum is 2.90 \AA, while the second 
one gives $a_0=2.88$ \AA.

\begin{figure}
\includegraphics[width=\columnwidth]{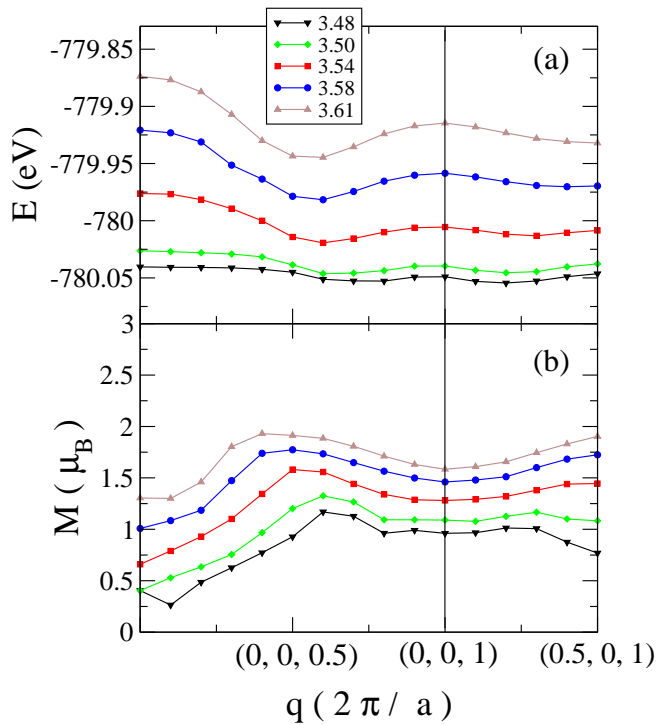}
\caption{\label{spiralLDA} (a) Free energy and (b) magnetic moment as a function of pitch
vector $\vec{q}$ of the spiral state for different lattice constants ranging from
3.48 \AA to 3.61 \AA.}
\end{figure}

We find that DZP BS predict erroneous orderings of the ground and first excited states. 
Such effect is particularly apparent in the LDA approximation. Fig. \ref{statesLDA} (a) shows that
BSD erroneously predicts that the LDA ferromagnetic bcc state is more stable that the
paramagnetic fcc one. Fig. \ref{statesLDA} (b) shows that usage of more complete BS correct 
such a mistake. Under such proviso, SIESTA provides pretty accurate results for the LDA 
predictions of the different physical magnitudes. For instance, the lattice constant,
magnetic moment and bulk modulus of ferromagnetic bcc Iron are found to be 2.76 \AA, 
$2.08 \mu_B$ and 2.68 Mbar, which compare extremely well with the best all-electrons 
Plane-wave calculations \cite{Wan85}. Moreover, the lattice constant for paramagnetic fcc, 
3.38 \AA, is also very similar to all-electrons estimate of 3.375 \AA, while the energy
difference between both states is somewhat underestimated (55 versus about 70-80 meV). 
\cite{Wan85}.     

\begin{figure}
\includegraphics[width=\columnwidth]{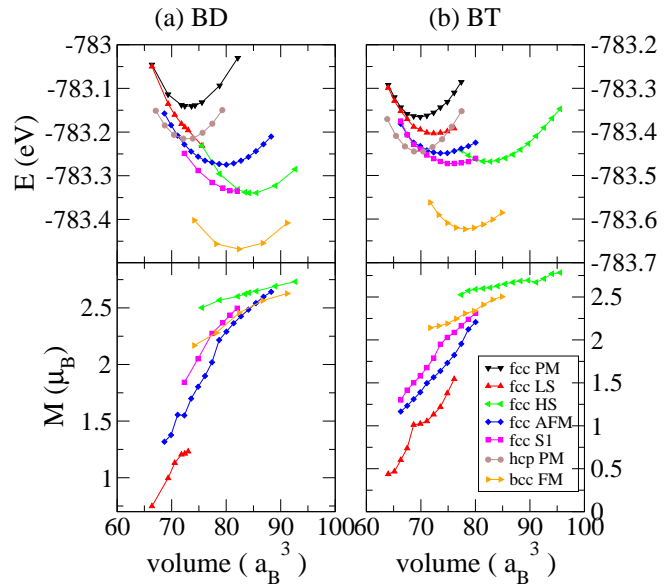}
\caption{\label{statesGGA} Free energy and magnetic moment of the ground and lowest excited
states of bulk Iron as predicted by LDA approximation, using (a) BSD and (b) BST, as
a function of the atomic volume.}
\end{figure}

DZP Basis sets also provide awkward results in the GGA approximation, even though the 
relative stability of the lowest energy states is correct now, see Fig. \ref{statesGGA}. 
Nevertheless, we find that the shape of the energy curves of the fcc states change
significantly when we increase the size of the basis from BSD to BST. Now, since the
spiral state smoothly interpolates between the ferromagnetic HS and the antiferromagnetic
ones, we have increased further the size of the BS. We have included more polarization 
orbitals of p, d and f symmetry, and have found that the energy of the three curves is 
essentially converged (see Fig. \ref{hilbert}). We find equilibrium lattice constant, 
magnetic moment and bulk modulus of 2.85 \AA, $2.31 \mu_B$ and 1.83 Mbar for the 
ground state, which compare reasonably well with former all-electrons or 
ultrasoft-pseudopotentials-based plane waves calculations \cite{Mor97,Kno00}. We have 
computed the properties of the spiral state for lattice constants well below 3.54 \AA,
so that the binding energy curve is a clear parabola, with a minimum at $a=3.56$ \AA,
very close to the experimental value (2.577 \AA). 

We have also simulated clusters with a number of atoms ranging from 2 to 5, using a BST and
non collinear GGA, as shown in Table \ref{Clusters}. Our calculations compare very well
with previous theoretical simulations \cite{Oda98,Hob00,Izq00,Die01,Pos03} and even improve
them when comparisons are made with the experimental values found for the $Fe_2$ cluster
\cite{Pur82}.

\begin{table}
\caption{Bond lengths $a$ (\AA), binding energy per atom $E_{b}$ (eV/atom)
and total magnetic moment $M$ ($\mu_{B}$) for Iron clusters up to 5 atoms calculated with a
Triple-Zeta basis and GGA.}
\label{Clusters}
\begin{tabular}{llll}
\hline\noalign{\smallskip}
& $a$ (\AA) & $B$ (eV/\AA) & $M$ ($\mu_{B})$\\
\noalign{\smallskip}\hline\noalign{\smallskip}
Fe$_{2}$ & 2.02 & 1.51 & 6.00\\
Fe$_{3}$ D$_{\infty h}$ & 2.28 & 1.72 & 5.62\\
Fe$_{3}$ C$_{3v}$ & 2.27 & 1.88 & 10.00\\
Fe$_{4}$ C$_{4v}$ & 2.30 & 2.21 & 14.00\\
Fe$_{4}$ T$_{d}$ &1,2$\leftrightarrow$3,4 2.27 & 2.31 & 14.00\\
& 1$\leftrightarrow$2, 3$\leftrightarrow$4 2.65 & & \\
Fe$_{5}$ D$_{3h}$ &1$\leftrightarrow$2,3 2$\leftrightarrow$3 2.43 & 2.58 & 17.07\\
& 1,2,3$\leftrightarrow$4,5 2.37 & & \\
\noalign{\smallskip}\hline
\end{tabular}
\end{table}

\subsection{Spiral states in the $\gamma$ phase}
We turn now to the predictions for the spiral state in LDA. We scan the
energy as function of pitch vector along the $\Gamma$X and XW directions, where we find 
the two minima $\vec{q}_1$ and $\vec{q}_2$ we talked about in the introduction. On closer 
inspection of Fig. \ref{spiralLDA}, we see that the energy curves for lattice constants
equal or larger than 3.58 \AA $\;$ only have the $\vec{q}_1$ minimum. The second minimum 
appears when we decrease $a$ at or below 3.54 \AA, becoming lowest in energy at
$a\approx3.50$ \AA. The curves corresponding to smallest lattice constants are very shallow.
Their two minima have almost the same energy up to 1 meV, and are separated by energy
barriers as small as 4 meV. If LDA were accurate enough for Iron, one would expect both
phases not only to coexist but also to change dynamically from one to
the other.

We finally discuss our results for the spiral structures in GGA, where
we find that the S2 state has already clearly developed when $a=3.52$ \AA, but that the
ground state is S1 down to lattice constants of 3.47 \AA. It can be seen again that the
energy curves change rather much when we increase the size of the basis. One of the 
reasons is that the shape of the binding energy parabola also changes significantly (see
Fig. \ref{statesGGA}). But for BST, the ferromagnetic state has always considerably higher
energy and there is a clearer asymmetric double-well structure with activation barriers
of about 5-7 meV.

Marsman and Hafner have also performed simulations of $\gamma$-Iron under tetrahedric,
orthorombic and monoclinic distortions. However they obtained for the undistorted case a
spiral state S2 with equilibrium lattice constant $a=3.49$ \AA, much smaller than the
experimental one. They also found that the equilibrium lattice constant for S1 was
$a=3.51$ \AA. On the contrary, we obtain a state S1 with $a=3.56$ \AA, much closer to
experiments.

\begin{figure}
\includegraphics[width=\columnwidth]{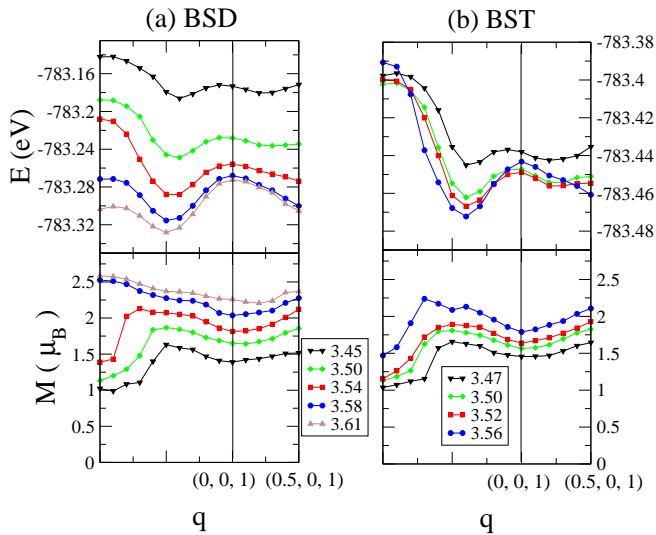}
\caption{\label{spiralGGA} (a) Free energy and (b) magnetic moment as a function of pitch
vector $\vec{q}$ of the spiral state for different lattice constants ranging from
3.48 \AA to 3.61 \AA.}
\end{figure}

\section{Summary}

We have performed a thorough study of the different Iron phases in order to provide
a good basis for future calculations. We have paid special attention to the minimization
of the different parameters used in the LCAO-DFT SIESTA code. We found that a grid
cutoff of 700 Ry, 4.000 k points and temperature smaller than 300 K are needed to meet
an accuracy of about 1 meV.

We showed that a double-zeta basis is not accurate enough to supply the correct ground
state ordering of the Iron phases, since for LDA it predicts that the bcc phase is less
stable than the fcc, which is contrary to all previous calculations. However, when we use
a triple-zeta basis the results change dramatically for both LDA and GGA.
We obtain for the ferromagnetic bcc and equilibrium lattice constant of
2.85 \AA, a magnetic moment of 2.31 $\mu_B$ and a bulk modulus of 1.83 Mbar, in excellent
agreement with experiments. We have also simulated Iron clusters and we found a better
estimate of the properties of these materials than previous works.

Finally, we have also made a profound study of the $\gamma$ phase and we found an equilibrium
lattice constant of about 3.56 \AA, closer to the experimental value of 3.577 \AA, but with
a spiral state S1 instead of the experimental S2. However these simulations agree with
previous works and even improve them.

\begin{acknowledgements}
J. F. wishes to thank to the authors and developers of the SIESTA code for a number of
helpful discussions and e-mail exchanges: P. Ordej\'on, J. Soler, E. Artacho, 
A. Garc\'{\i}a-Arribas and J. Junquera. He also wishes to acknowledge discussions with 
A. Vega and S. Bouarab. Optimization of the BS was carried out with the use of the 
package INNERGRID which was generously supplied by GRID SYSTEMS. The work presented here 
has been funded by the Spanish MCyT and the European Union under project no.
BFM2000-0526 and contract no. HPRN-CT-2000-00144, respectively. V. M. G.-S. thanks the Ministerio
Espa\~nol de Educaci\'on, Cultura y Deporte for a fellowship.
\end{acknowledgements}


\begin{thebibliography}{31}

\bibitem{Koh65} W. Kohn, and L. J. Sham, Phys. Rev. \textbf{140}, A1133 (1965).

\bibitem{Sol02} J. M. Soler et al., Journal of Physics Condensed Matter \textbf{14},
2745 (2002).

\bibitem{Tro91} N. Troullier and J. L. Martins, Phys, Rev. B \textbf{43}, 1993 (1991).

\bibitem{Jun01} J. Junquera, O. Paz, D. S\'anchez-Portal, and E. Artacho, Phys. Rev. B
\textbf{64}, 235111 (2001).

\bibitem{Ang02} E. Anglada, J. M. Soler, J. Junquera, and E. Artacho, Phys. Rev. B
\textbf{66}, 205101 (2002).

\bibitem{Kno00} K. Kn\"opfle, L. M. Sandratskii, and J. K\"ubler, Phys. Rev. B
\textbf{62}, 5564 (2000).

\bibitem{Kub88} J. K\"ubler, K. H. Hock, J. Sticht, and A. R. Williams, J. Phys. F. Met. Phys.
\textbf{18}, 469 (1988).

\bibitem{San98} L. M. Sandratskii, Adv. Phys. \textbf{47}, 91 (1998).

\bibitem{Sti89} J. Sticht, K. H. Hock, and J. K\"ubler, J. Phys.: Condens. Matter
\textbf{1}, 8155 (1999).

\bibitem{Spi02} D. Spisak and J. Hafner, Phys. Rev. Lett. \textbf{88},
056101 (2002).

\bibitem{Tsu89} Y. Tsunoda, J. Phys.: Condens. Matter \textbf{1}, 10427 (1989).

\bibitem{Tsu93} Y. Tsunoda, Y. Nishioka, and R. M. Nicklow, J. Magn. Magn. Materials
\textbf{128}, 133 (1993).

\bibitem{Uhl94} M. Uhl, L. M. Sandratskii, and J. K\"ubler, Phys. Rev. B \textbf{50},
291 (1994).

\bibitem{Mor97} E. G. Moroni, G. Kresse, J. Hafner, and J. Furtmuller, Phys. Rev. B
\textbf{56}, 15629 (1997).

\bibitem{Kor96} M. Korling and J. Ergon, Phys. Rev. B \textbf{54}, R8293 (1996).

\bibitem{Mar02} M. Marsman and J. Hafner, Phys. Rev. B \textbf{66}, 224409 (2002).

\bibitem{Uhl92} M. Uhl, L. M. Sandratskii, and J. K\"ubler, J. Magn. Magn. Materials
\textbf{103}, 314 (1992).

\bibitem{Byl98} D. M. Bylander and L. Kleinman, Phys. Rev. B \textbf{58},
9207 (1998).

\bibitem{Byl99a} D. M. Bylander and L. Kleinman, Phys. Rev. B \textbf{59},
6278 (1999).

\bibitem{Byl99b} D. M. Bylander and L. Kleinman, Phys. Rev. B \textbf{60},
R9916 (1999).

\bibitem{Dzy58} I. Dzyaloshinskii, J. Phys. Chem. Solids \textbf{4},
241 (1958).

\bibitem{Mor60} T. Moriya, Phys. Rev. \textbf{120}, 91 (1960).

\bibitem{Kle99} L. Kleinman, Phys. Rev. B \textbf{59}, 3314 (1999).

\bibitem{Izq00} J. Izquierdo et al., Phys. Rev. B \textbf{61}, 13639 (2000).

\bibitem{Mor92} J. Moreno and J. Soler, Phys. Rev. B \textbf{45}, 13891 (1992).

\bibitem{Per96} J. Perdew, K. Burke, and M. Ernzerhof, Phys. Rev. Lett. \textbf{77},
3865 (1996).

\bibitem{Wan85} C. S. Wang, B. M. Klein, and H. Krakauer, Phys. Rev. Lett.
\textbf{54}, 1852 (1985).

\bibitem{Oda98} T. Oda, A. Pasquarello, and R. Car, Phys. Rev. Lett. \textbf{80},
3622 (1998).

\bibitem{Die01} O. D. Dieguez et al., Phys. Rev. B \textbf{63}, 205407 (2001).

\bibitem{Pos03} A. V. Postnikov, P. Entel, and J. M. Soler, submitted to Eur. Phys. J.

\bibitem{Hob00} D. Hobbs, G. Kresse, and J. Hafner, Phys. Rev. B \textbf{62},
11556 (2000).

\bibitem{Pur82} H. Purdum, P. A. Montano, G. K. Shenoy, and T. Morrison,
Phys. Rev. B \textbf{25}, 4412 (1982).

\end{thebibliography}
\end{document}